\documentclass[%
reprint,
amsmath,amssymb,
aps,prl,nobibnotes
]{revtex4-2}

\usepackage{graphicx}% Include figure files
\usepackage{dcolumn}% Align table columns on decimal point
\usepackage{bm}% bold math
\usepackage[dvipsnames]{xcolor}
\begin{document}

\preprint{APS}

\title{Optical Feedback Loop in Paraxial Fluids of Light: A Gate to New Phenomena in Analogue Physical Simulations}
% Force line breaks with \\

\author{Tiago D. Ferreira}
\thanks{Authors contributed equally}
\email{tiago.d.ferreira@inesctec.pt}
\author{Ariel Guerreiro}
\author{Nuno A. Silva}
\thanks{Authors contributed equally}
\email{nuno.a.silva@inesctec.pt}
\address{Departamento de Física e Astronomia, Faculdade de Ciências, Universidade do Porto, Rua do Campo Alegre s/n, 4169-007 Porto, Portugal}
\address{INESC TEC, Centre of Applied Photonics, Rua do Campo Alegre 687, 4169-007 Porto, Portugal}

\date{\today}% It is always \today, today,

\begin{abstract}
Easily accessible through tabletop experiments based on laser propagation inside nonlinear optical media, Paraxial Fluids of Light are emerging as promising platforms for the simulation and exploration of quantum-like phenomena. In particular, the analogy builds on a formal equivalence between the governing model for a Bose-Einstein Condensate under the mean-field approximation and the model of laser propagation under the paraxial approximation. Yet, the fact that the role of time is played by the propagation distance in the optical analogue system may impose strong bounds on the range of accessible phenomena due to the limited length of the nonlinear medium. In this manuscript, we present a novel experimental approach to solve this limitation in the form of an optical feedback loop, which consists of the reconstruction of the optical states at the end of the system followed by their subsequent re-injection exploiting wavefront shaping techniques. The results enclosed demonstrate the potential of this approach to access unprecedented dynamics, paving for the observation of novel phenomena in these systems.
 
\end{abstract}

\maketitle

The subject of analogue physical simulators with optical systems has grown significantly in recent years, transitioning from a period of theoretical and numerical predictions to a stage where tabletop experiments are increasingly prevalent \cite{ferreira2022towards,ferreira2023exploring,boughdad2020fluids,abobaker2022inverse,azam2022vortex}. In particular, one of the hallmarks of the subject is the emulation of quantum fluids, relying on the universal character of the nonlinear Schrodinger equation (NSE). Indeed, as the NSE model describes both the dynamics of a Bose-Einstein Condensate(BEC) under the mean-field approximation \cite{pitaevskii2016bose} as well as the dynamics of the electromagnetic field in a nonlinear optical medium either in propagating \cite{carusotto2014superfluid,larre2015propagation} or confined geometries \cite{carusotto2013quantum}, it enables analogue physical simulations of quantum fluids phenomenology through a formal mathematical equivalence. Yet, while the versatility of these experimental systems is typically much wider than a BEC, these analogue platforms still display important limitations for the deployment of an effective quantum fluids simulator.

Although propagating (also called paraxial fluids of light) and confined (often called polariton quantum fluids) geometries are both fascinating platforms for the emulation of quantum-like phenomenology, their differences lead to very specific advantages. First, polariton fluids have an underlying quantum phenomenology governing the dynamics, allowing to access truthful quantum dynamics and applications \cite{dominici2018interactions}, contrary to paraxial fluids, whose behavior is completely classic. Yet, this comes at the cost of a driven-dissipative regime \cite{carusotto2013quantum,ferdinand_high_2022}, which not only hinder the range of phenomena that can be emulated with such systems but may also require cryogenic temperatures \cite{ferdinand_high_2022}, possibly augmenting the complexity of the experimental platform. Taking this context, paraxial fluids of light appear as a promising toolbox for the physical simulation of quantum-like effects, featuring as major advantages being easier to model and realize experimentally \cite{carusotto2014superfluid,abobaker2022inverse,braidotti2022measurement}, being easier to manipulate and assess with wavefront shaping and holography techniques \cite{ferreira2022towards,OADH_method}, and presenting lower optical power losses \cite{ferreira2022towards}.

In general concepts, the experimental emulation of quantum-like phenomenology with paraxial fluids of light explores the mathematical equivalence between the Gross-Pitaevskii mean-field description of a BEC and the paraxial approximation of the propagation of a continuous-wave optical beam in a nonlinear medium \cite{carusotto2014superfluid,larre2015propagation}. In this analogy, the electric field intensity acts as a fluid density and the propagation direction plays the role of an effective time \cite{carusotto2014superfluid,larre2015propagation}. In recent years, a plethora of quantum-like effects was experimentally observed, including the superfluidity of light and drag-force suppression using induced all-optical defects \cite{ferreira2022towards,michel2018superfluid}, the study and measurement of a Bogoliubov dispersion relation for the elementary excitations on top of a constant density fluid \cite{piekarski2021measurement,fontaine2018observation}, the observation of a two-dimensional vortex turbulence regime \cite{eloy2021experimental} and its characteristic power laws \cite{ferreira2022towards,abobaker2022inverse}, and even analogue cosmological particle creation \cite{steinhauer2022analogue} and superradiance-like signatures \cite{braidotti2022measurement}.  Concerning the nonlinear media utilized, different optical materials have been explored in the literature, with hot atomic vapors \cite{glorieux2023hot} and photorefractive crystals being two of the most successful research lines \cite{ferreira2022towards,boughdad2020fluids}. 

Yet, in spite of the wide versatility of paraxial fluids of light, they still present some drawbacks that can strongly limit their range of applicability, being the most important the optical absorption and the limited size of the propagation length. The first one is inherent to all the nonlinear optical mediums and may be mitigated depending on the choice of media and conditions utilized (e.g. typically lower in photorefractive crystals than in hot atomic vapors). The second one, concerns the physical dimensions of the nonlinear medium (e.g. size of the crystal in the propagation direction) and for practical applications can be far more limitative and harder to circumvent. Indeed, since the propagation distance in these fluids of light is mapped onto an effective time, the limited size of the nonlinear mediums bounds the maximum simulation time that we can observe experimentally. A way to circumvent this limitation is to artificially control the effective length of these systems by varying the intensity of the beam and consequently the nonlinear response of the system which effectively rescales the governing equation \cite{abuzarli2021blast,abobaker2022inverse,ferreira2023exploring}. Yet, such a methodology requires a rescaling of the transverse spatial dimensions of the input beam, which may be challenging to achieve. Besides, concerning the increase of the propagation distance, it is associated with the increase of the optical power, typically leading to the saturation of the optical media \cite{bienaime2021quantitative}. For this specific challenge, some of the media may be suitable for physical extensions of the propagation length, but such alternatives may become expensive or limited by absorption.

In this letter, we focus on this major limitation, proposing an experimental technique based on an optical reinjection loop - the so-called feedback loop - that utilizes trustworthy amplitude-phase reconstruction using digital off-axis holography \cite{OADH_method} and convenient wavefront shaping techniques. Through significant case studies we demonstrate that our experimental methodology opens a gate to assess unprecedented dynamical regimes and introduces new degrees of versatility featuring non-trivial opportunities in terms of the possible experimental scenarios.

%%%%%%%%%%%%%%%%%%%%%%%%%%%%%%%%%%%%%%%%%%%
%%%%%%%%%%%%%%%%%%%%%%%%%%%%%%%%%%%%%%%%%%%
%%%%%%%%%%%%%%%%%%%%%%%%%%%%%%%%%%%%%%%%%%%

\emph{\textbf{Physical Model - }} To model a paraxial light fluid, we assume a continuous wave laser propagating along the axis $z$ of a nonlinear medium expressed as $\boldsymbol{E}_f(\boldsymbol{r}_\perp, z)=E_f(\boldsymbol{r}_\perp,z)\exp{[i (n_e k_f z-\omega t)]}\boldsymbol{e}_p + c.c.$, with $E_f(\boldsymbol{r}_\perp,z)$ being the envelope function, $k_f=2\pi /\lambda_f$ being the vacuum wavenumber. Considering the specific choice of polarization $\boldsymbol{e}_p$ aligned with the $c-$axis of the photorefractive crystal, and neglecting the transient and anisotropic response of the crystal, the propagation dynamics of the spatial envelope of the optical beam may be described under the paraxial approximation as \cite{boughdad2020fluids, ferreira2022towards}
\begin{multline}
i \frac{\partial E_f}{\partial z} + \frac{1}{2n_ek_f} \nabla_\perp^2 E_f - \\k_f\Delta n_{max} \frac{|E_f|^2}{|E_f|^2+ I_{sat}} E_f +i\frac{\alpha}{2}E_f= 0,
\label{eq:fluid_eq}
\end{multline}
where $\alpha$ is the medium absorption, $I_{sat}$ is a  saturation intensity that may be experimentally controlled using an incoherent white light, and $\Delta n_{max}=1/2n_e^3r_{33}E_{ext}$ a maximum nonlinear refractive index variation. For the purpose of this work, we utilized $\lambda_{f}=532nm$ propagating in a $5mm\times5mm\times20mm$ SBN:61 crystal (Ce doped at $0.002 \%$) biased with an external static electric field  $E_{ext} = 8\times10^{4}V/m$ and white light intensity $I_{sat} = 200mW/cm^2$. Following the experimental procedure outlined in ref. \cite{boughdad2019anisotropic}, we obtain $\Delta n_{max}\approx 1.25\times10^{-4}$ and $\alpha\approx0.167cm^{-1}$ for such conditions.

\begin{figure}
\begin{center}
\includegraphics[width=0.5\textwidth]{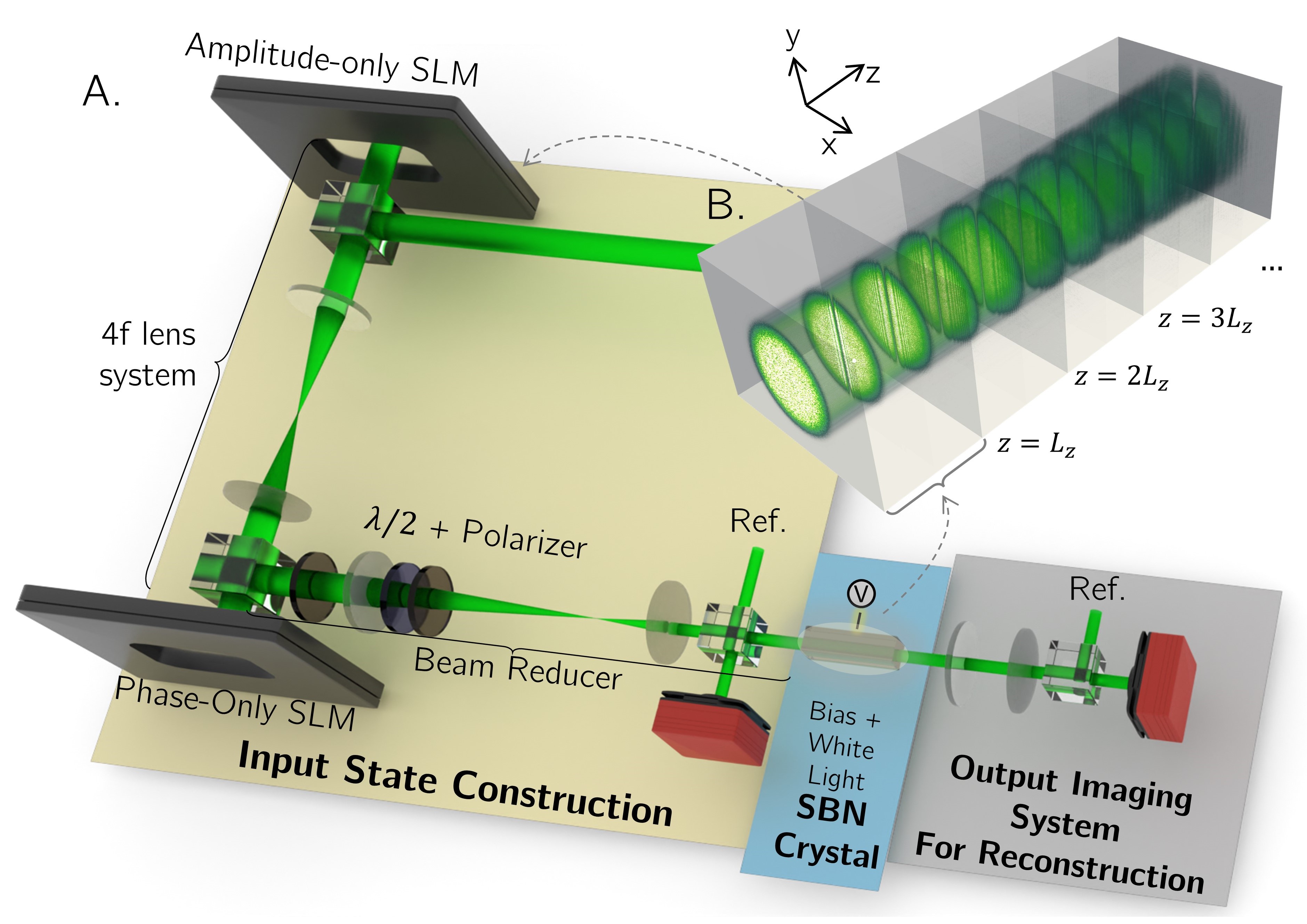}
\end{center}
\caption{Simplified Illustration of the experimental setup, exploiting an SLM as an amplitude modulator matched at the single-pixel level to a second SLM acting as a phase-only modulator to achieve full control of the input state. After the propagation inside the nonlinear crystal, the reference arm is used for the reconstruction of the optical beam at the output face of the crystal using a digital off-axis holography technique. This way it is possible to extend the effective propagation distance of our nonlinear medium and thus increase the analogue simulation time of the paraxial fluid of light.}
\label{fig:experimental_setup}
\end{figure}

To better establish the connection with the literature related to BECs and for comparison purposes regarding the maximum propagation distance, one can now transform the NSE into an adimensional form by focusing on the linear regime $|E_f| \ll I_{sat}$. For that, we first take the scalings $\boldsymbol{r}_{\perp}^{\prime}=k_f \sqrt{n_e \Delta n}\boldsymbol{r}_{\perp}$, $\tau=k_f \Delta n$ with $\Delta n = \Delta n_{max} I_f^0/I_{sat}$ where $I_f^0$ is defined via the transformation $E_f = \sqrt{I_f^0} E^{\prime}_{f}$ with $|E^{\prime}_f|$ being a normalized envelope between 0 and 1. Dropping the primes for simplicity, yields
\begin{equation}
i  \frac{\partial E_f}{\partial \tau} + \frac{1}{2} \nabla_\perp^2 E_f - |E_f|^2 E_f +i\frac{\alpha^{\prime}}{2} E_f= 0,
\label{eq:fluid_eq_ad}
\end{equation}
with $\alpha^{\prime}=\alpha/(k_f\Delta n)$.
Put in this form, it is easier to provide estimates for the maximum propagation distance/effective time $\tau_{max} = k_f \Delta n L_z$ for distinct nonlinear media explored in the literature, where $L_z$ equals the length of the media. Typical values for $\tau_{max}$ lay on the range \emph{[0-15]} for photorefractive crystals \cite{ferreira2022towards,ferreira2023exploring,michel2018superfluid}, \emph{[0,200]} for hot atomic vapors, but at a cost of a significantly higher absorption coefficient, $\alpha\approx0.78cm^{-1}$ \cite{abobaker2022inverse}, and \emph{[0,15]} for thermo-optic materials \cite{vocke2015experimental}.

\emph{\textbf{Experimental Setup and the Optical Feedback Loop - }}  The envelope electric field can be expressed as $E_f(x,y)=A(x,y)e^{i\phi(x,y)}$, where real functions $A(x,y) = \sqrt {I_f (x,y)}$ and $\phi(x,y)$ relate to the beam intensity and phase profile, respectively. Taking this into consideration, and inspired by concepts previously explored in Ref. \cite{sun2012observation}, we may achieve full control of the input state by using two Spatial Light Modulators (SLMs) in a perpendicular configuration as illustrated in Figure \ref{fig:experimental_setup}, where the first one modulates the optical beam in amplitude, whereas the second modulates the wavefront in phase.

To achieve the optical feedback loop, the output plane of the crystal is imaged with an output camera using a 4-f lens system, being the phase reconstructed with a digital holography method using an unmodulated external reference \cite{OADH_method}. After reconstruction of the output beam, the intensity and phase profiles are interpolated to the resolution and pixel pitch of the SLMs screen and imprinted on the amplitude and phase-only SLMs respectively (see Figure \ref{fig:experimental_setup}, and \cite{supp} for a complete scheme and additional details). For this process, the screen of the first SLM is imaged at the screen of the second SLM using a 4-f lens system, matching both phase and amplitude masks with an error of the order of the SLMs pixel size by utilizing linear translation stages. Subsequently, an additional beam reducer forms the image plane of the SLMs at the input face of the crystal. Finally, the total power of the generated optical beam is adjusted to match the one measured at the output utilizing a combination of a half-wave plate and a polarizing beam-splitter, completing the output state reconstruction.

\begin{figure}
\begin{center}
\includegraphics[width=0.5\textwidth]{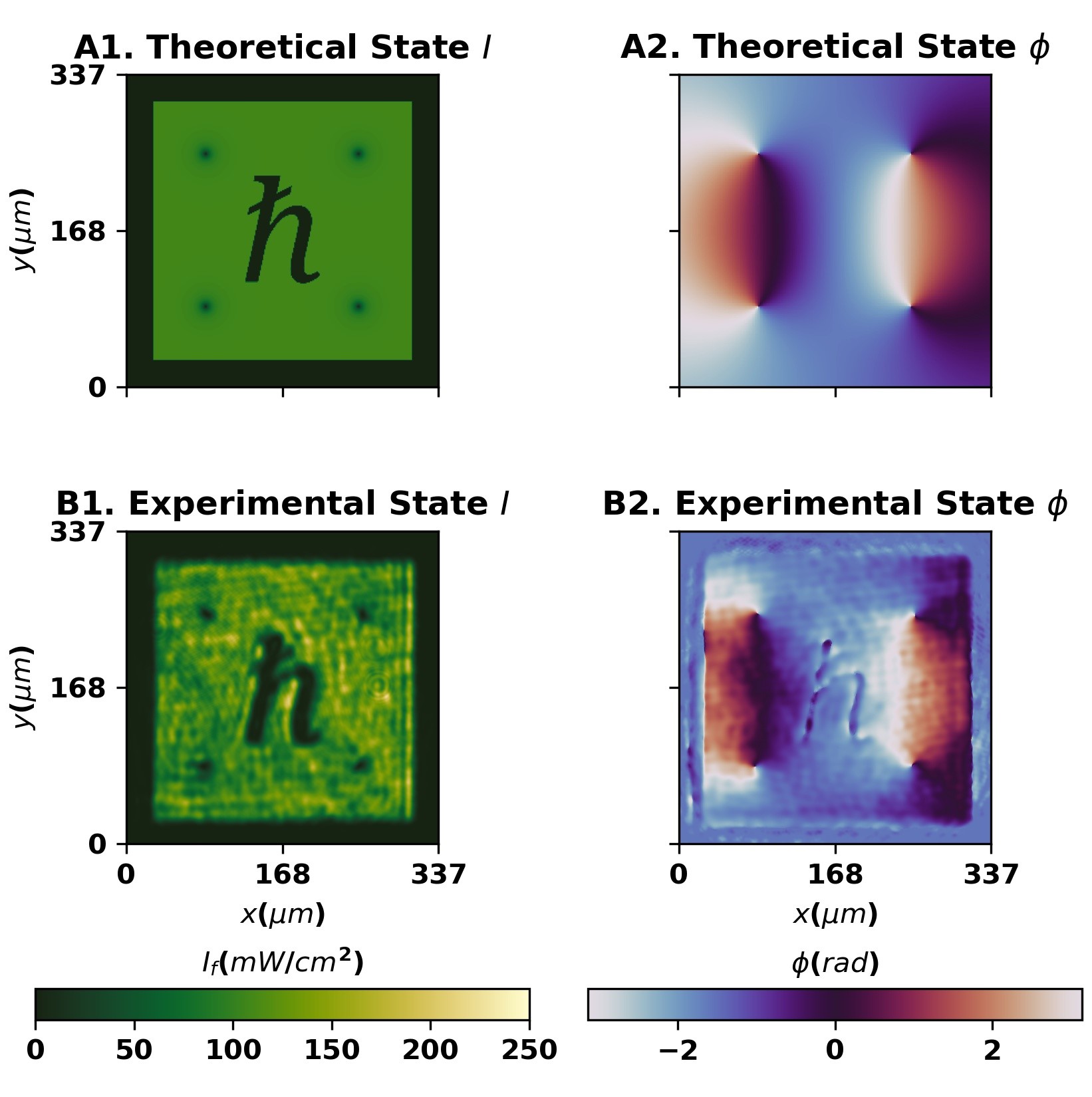}
\end{center}
\caption{Arbitrary state generation. In subfigures A we have the numerical state, and in B the measured experimental state. The results were obtained at the crystal input image plane.}
\label{fig:arbitrary_state_generation}
\end{figure}

In order to validate this solution, we reconstructed the state at the input of the crystal obtaining $A_{in,exp}$ and $\phi_{in,exp}$ to monitor and evaluate the reconstruction quality of both amplitude and phase of the input state in comparison to a target $A_{in}$ and $\phi_{in}$. Taking an arbitrary shape as the case illustrated in Figure \ref{fig:arbitrary_state_generation}, one can qualitatively observe that both amplitude and phase are being correctly constructed at the input plane of the crystal. Furthermore, constructing a quantitative metric based on amplitude reconstruction at the input camera  for each spatial point $(x,y)$ as the Absolute Percentage Error, i.e.
\begin{equation}
APE (|A_{in,exp}(x,y)|) = 100 \times \left| 1- \frac{|A_{in,exp}(x,y)|}{|A_{in}(x,y)|}\right|\%,
\label{eq:MAPE_amp}
\end{equation}
we can obtain a mean absolute percentage error of 6.44\% for the reconstructed state in Figure \ref{fig:arbitrary_state_generation}. As discussed in more detail in the supplementary material \cite{supp}, the error may be recreated with a random Gaussian noise applied at each passage, whose properties will be used for the initial state of the numerical simulations. Besides, as the noise is only added at the input of each passage, we do not expect major effects to occur on the overall qualitative behavior of the simulation. Yet, and as further detailed in the supplementary material \cite{supp}, in some scenarios of lower signal-to-noise ratio or involving instability analysis, this additional noise shall be taken into consideration for quantitative analysis as it may hinder observations or introduce unwanted artifacts.

With the construction and reconstruction stage validated, the experimental configuration proposed brings two non-trivial advantages to the subject of analogue simulations with paraxial fluids of light. On one hand, this configuration enables the re-injection of the output state, effectively extending the maximum propagation distance $\tau_{max}$ accessible. On the other hand, allowing full control of the initial input state, this configuration presents an unprecedented degree of versatility that cannot be achieved with previous setups, allowing one to easily recreate complex experimental conditions.

%%%%%%%%%%%%%%%%%%%%%%%%%%%%%%%%%%%%%%%%%%%%%%%%%%%%%%%%%%%%%%%
%%%%%%%%%%%%%%%%%%%%%%%%%%%%%%%%%%%%%%%%%%%%%%%%%%%%%%%%%%%%%%%
%%%%%%%%%%%%%%%%%%%%%%%%%%%%%%%%%%%%%%%%%%%%%%%%%%%%%%%%%%%%%%%

\emph{\textbf{Shock Waves and Dark Soliton Decay - }} To validate the capacity of our experimental solution, we studied a well-known scenario involving the formation of a dispersive shock-wave and snake instability of a dark soliton due to an initial phase slip. This dynamical behavior is extensively explored in the literature, mostly in numeric manners in the context of BECs \cite{ohya2019decay}, with a previous approach with nonlinear optics in the paraxial approximation presented in Ref. \cite{DS_atoms,DS_photorefractive,azam2022vortex} and in polariton fluids in Ref. \cite{claude2020taming}. For this purpose, the initial state is a modulated supergaussian 
\begin{equation}
    E_f(x,y)=\sqrt{I_0}e^{-\frac{1}{2}\left(2\frac{x^2+y^2}{w_e^2}\right)^4}e^{i\pi H(x)},
\end{equation}
where $w_e\approx650\mu m$, $I_0=40mW/cm^2$, and $H(x)$ is the Heaviside function to imprint a phase slip at $x=0$.

\begin{figure}
\begin{center}
\includegraphics[width=0.5\textwidth]{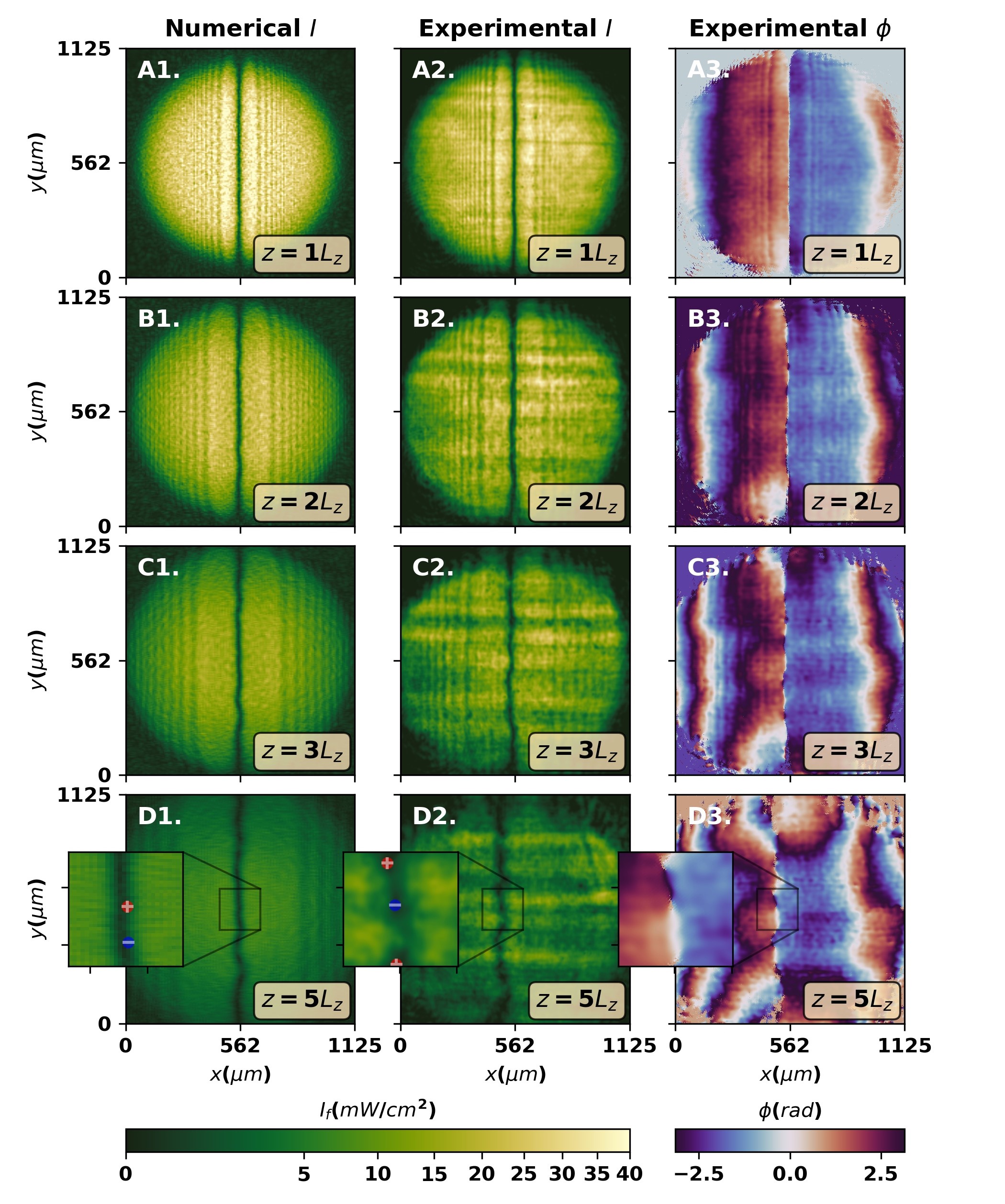}
\end{center}
\caption{Dark soliton decay, snake instability, and shock waves expansion. The first column corresponds to the numerical simulations, while the last two to experimental data, with the amplitude and phase profiles, corresponding to the middle and last columns, respectively. In the last row, a section of the system is highlighted to show the dark soliton decay into vortex pairs. The red circles with a "$+$" and the blue circles with a "$-$" correspond to a vortex of positive and negative circulation, respectively. See \cite{supp} for all the passages.}
\label{fig:dark_soliton_decay}
\end{figure}

\begin{figure}
\begin{center}
\includegraphics[width=0.5\textwidth]{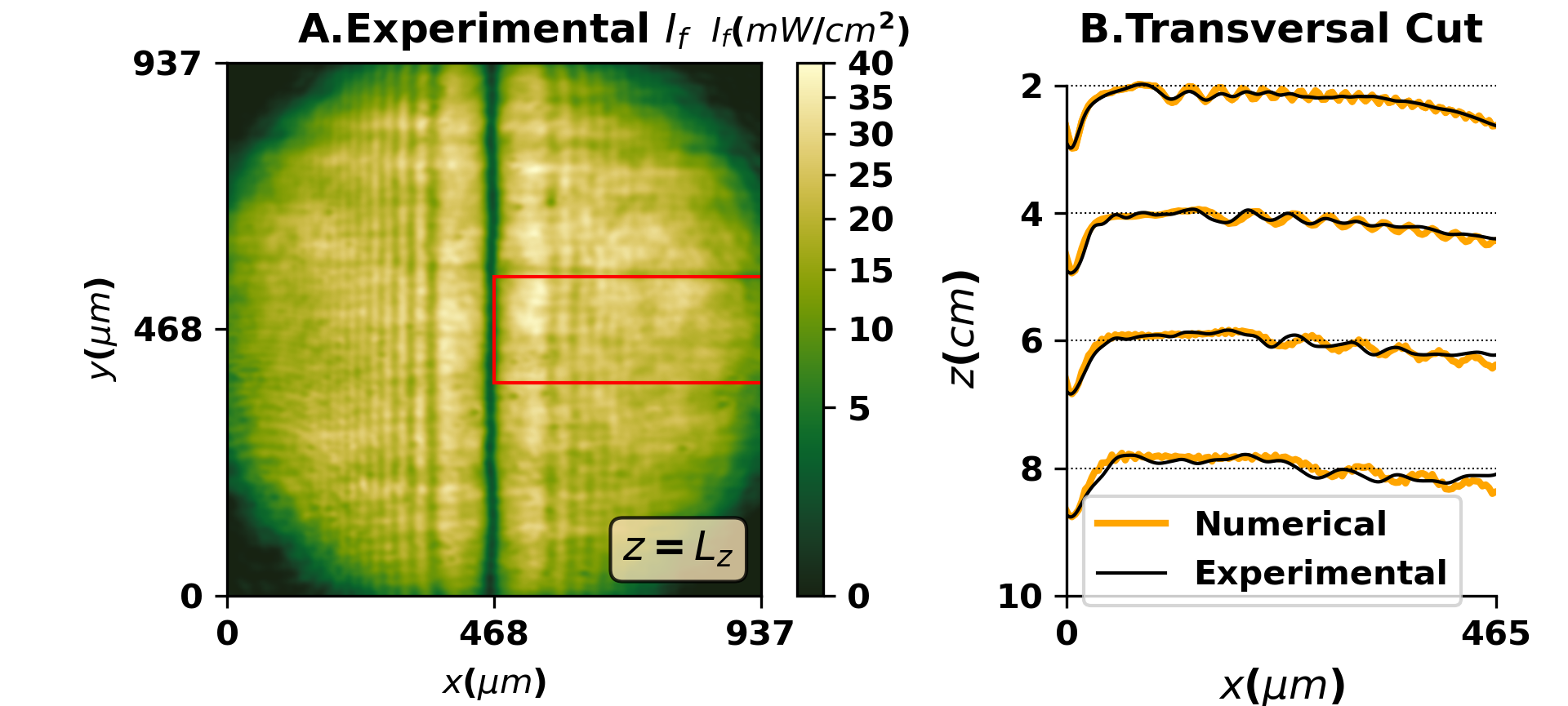}
\end{center}
\caption{Profile comparison between experimental and numerical results of the shock-wave propagation for the dark soliton decay in Figure \ref{fig:dark_soliton_decay}. Figure A., intensity profile after the first passage, highlighting the area that we use to perform a transversal cut on the $x$ axis, averaged along the $y-$axis. Figure B. shows the results for this transversal cut, comparing experimental and numerical results.}
\label{fig:darksol_sup_speed_of_sound}
\end{figure}

Utilizing the optical feedback loop, we perform 5 experimental passages inside the photorefractive crystal, effectively increasing the $\tau_{max}$ from $\tau_{max}\approx5.8$ to $\tau_{max}\approx29.0$. The experimental data obtained is illustrated in Figure \ref{fig:dark_soliton_decay} and Figure \ref{fig:darksol_sup_speed_of_sound}, alongside numerical results from a standard beam propagation simulation of equation \ref{eq:fluid_eq} \cite{ferreira2022towards} with similar initial conditions. Qualitatively, it is observed that both numerical and experimental results agree quite well: the phase slip first generates a dispersive shock-wave \cite{ivanov2020formation}, created in the process a dark soliton localized at the interface $x=0$.

Due to the instability of the dark soliton solution in two transverse dimensions, the existing noise is amplified, subsequently turning into a snake instability pattern that leads to the decay of the dark soliton into vortex pairs. Again, looking at panels C-D of Figure \ref{fig:dark_soliton_decay}, this expected behavior is present both in the numerical and experimental results, validating the solution. We shall stress however that due to the fact that the decay process is associated with the instability of the dark soliton solution, the outcome of both the experiment and numerical results strongly depends on the noise level and its distribution, meaning that achieving the same vortex positions in numerical simulations as in experimental results is unlikely due to the sensitivity of this decay process to the initial conditions. Furthermore, while we have assumed an isotropic response for the sake of simplicity, the effective anisotropic response of the crystal \cite{boughdad2019anisotropic} may deform the velocity field of the vortices and break their stability \cite{mamaev1996vortex,mamaev1997time}. From a broader perspective, this can impose restrictions on the number of passages that we can obtain with our system. Nevertheless, even with such restrictions the qualitative and quantitative match with the experimental results, Figures \ref{fig:dark_soliton_decay} and \ref{fig:darksol_sup_speed_of_sound}, respectively, validate the optical feedback loop, demonstrating that we can indeed access dynamics and track properties of the system evolution that were previously unattainable within our experiment. A complementary analysis of these results is found in the supplementary material \cite{supp}.

%We also calculated the incompressible kinetic energy spectrum for each passage, and we can see the appearance of the $k^{-3}$ power law after the end of the third passage, where the phase slip finally decays into vortex pairs.
%A problem that arises in this configuration and has not yet been solved is the interaction quenches that happen at the input and output of the crystal. When these quenches occur, especially at the input, Bogoliubov modes are generated in both opposite directions. This mode generation makes the output states no longer solutions of the nonlinear system, and so each passage is an approximation of the expected solutions. In any case, the obtained results are qualitatively similar to the numerical predictions, and so we can probe dynamics was impossible with previous experiment refartigo.  The effect of the quench is very visible at the end of the second passage, where we see some interference's in the middle. To mitigate this effect we add some random noise before imprinting the amplitude profile for the next passage. This noise will generate a large amount of Bogoliubov modes in all directions which, in mean, will cancel each other out. 
%%%%%%%%%%%%%%%%%%%%%%%%%%%%%%%%%%%%%%%%%%%%%%%%%%%%%%%%%%%%%%%
%%%%%%%%%%%%%%%%%%%%%%%%%%%%%%%%%%%%%%%%%%%%%%%%%%%%%%%%%%%%%%%
%%%%%%%%%%%%%%%%%%%%%%%%%%%%%%%%%%%%%%%%%%%%%%%%%%%%%%%%%%%%%%%

\begin{figure}
\begin{center}
\includegraphics[width=0.5\textwidth]{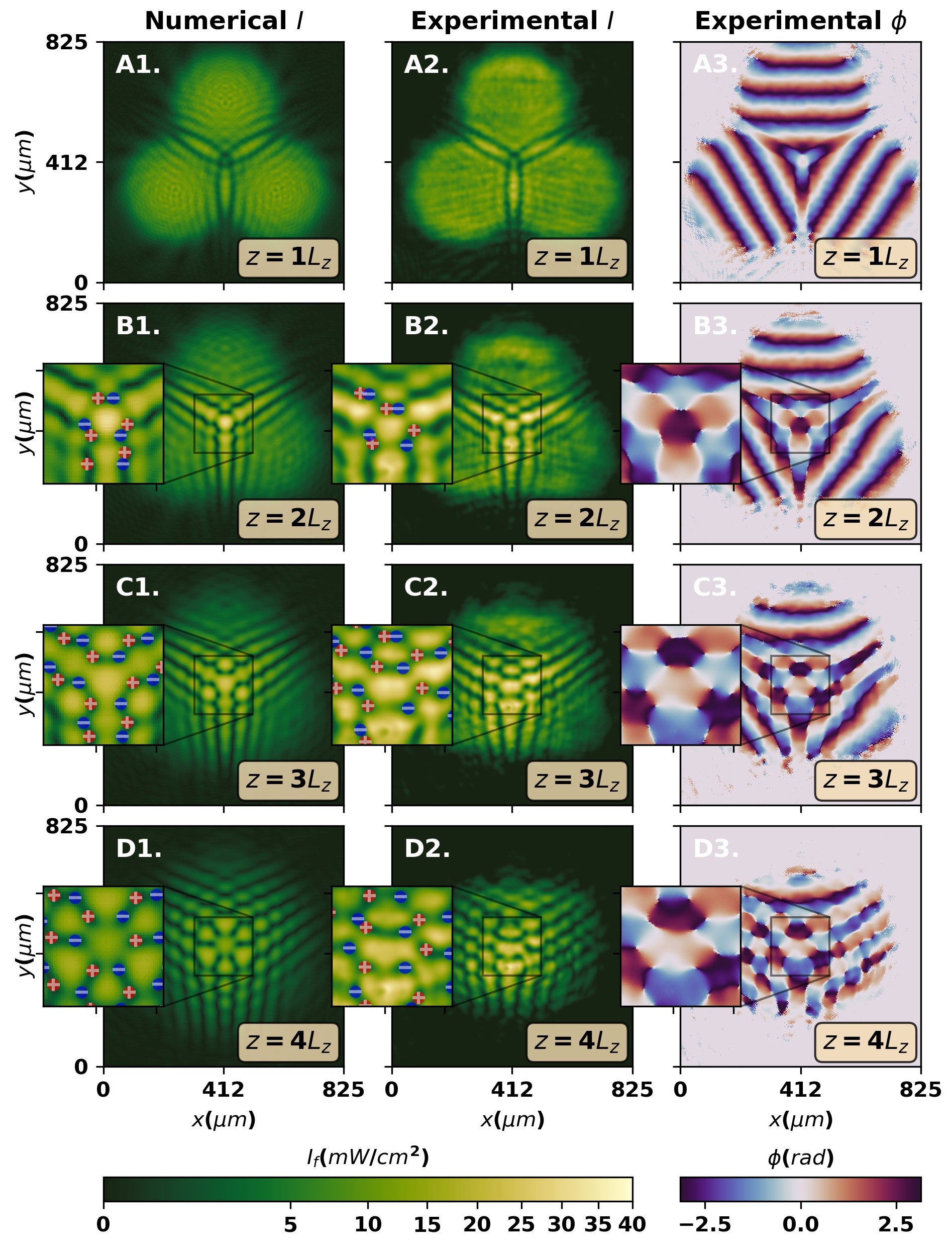}
\end{center}
\caption{Collision dynamics between three flat-top states. The first column corresponds to the numerical simulations, while the last two to experimental data, with the amplitude and phase profiles, corresponding to the middle and last columns, respectively. The center of the collision is highlighted to help the comparison between the numerical and experimental data. The red circles with a "$+$" and the blue circles with a "$-$" correspond to the vortices of positive and negative circulation, respectively. See \cite{supp} for all the passages.}
\label{fig:collision}
\end{figure}

\emph{\textbf{Flat-top collision - }} After validating the experimental configuration to extend the available range of the effective simulation, we now look at the potential of generating arbitrary configurations. For this, inspired by previous works \cite{abobaker2022inverse,boulier2016lattices} in related fields, we explored the dynamics of the collision of three supergaussian states
\begin{equation}
    E_f(x,y)=\sum_{j=1}^{j=3}\sqrt{I_0}e^{-\frac{1}{2}\left(2\frac{(x-x_j)^2+(y-y_j)^2}{w_e^2}\right)^4}e^{i\phi_j},
\end{equation}
where $w_e\approx280\mu m$, and $I_0 = 15mW/cm^2$. The phase distribution for each state is given by $\phi _j = v\left[ (x-x_j) cos(\theta _j) + (y-y_j) sin(\theta_j)\right]$, with $v = 4/6 \times 10^5 m^{-1}$, and $\theta_j = \pi/2+j2\pi/3$. The center positions are $(x_j,y_j)=R(cos(\theta _j) ,sin(\theta _j) )$ with $R=242\mu m$. The experimental results and numerical simulations are presented in Figure \ref{fig:collision}. Again, both numerical and experimental results show a good qualitative agreement, especially focusing on the polygonal structure of multiple vortices that are formed near the center. In particular, we highlight the fact that it is possible to track the vortices dynamics and interactions (see additional results in supplementary material \cite{supp}), which may pave for innovative studies on vortex dynamics and turbulent regimes \cite{kwon2021sound, panico2023onset}. 

As in the previous case, the anisotropic response of the crystal seems to play an increasingly relevant role with the accumulation of multiple passages and eventually translate into observable effects, in particular breaking the symmetry of the central structure. Yet, we also notice that contrary to the previous case study, the vortices in this configuration are less affected by the anisotropy of the crystal. This happens because the system now has many vortex pairs close to each other and with different orientations concerning the $c-$axis, which are known to be more stable \cite{mamaev1996vortex,mamaev1997time}. 

%Besides, this vortex lattice can be controlled through the initial velocity, which may be interesting for future studies.

%%%%%%%%%%%%%%%%%%%%%%%%%%%%%%%%%%%%%%%%%%%%%%%%%%%%%%%%%%%%%%%
%%%%%%%%%%%%%%%%%%%%%%%%%%%%%%%%%%%%%%%%%%%%%%%%%%%%%%%%%%%%%%%
%%%%%%%%%%%%%%%%%%%%%%%%%%%%%%%%%%%%%%%%%%%%%%%%%%%%%%%%%%%%%%%

\emph{\textbf{Concluding remarks - }} In this work, we proposed an optical feedback loop experimental methodology to circumvent the limited effective time of analogue quantum simulations with paraxial fluids of light due to the finite length of the nonlinear media. The feedback loop method consists on measuring the state at the exit of the nonlinear medium, reconstructing it back again at the input, in such a way that it is possible to propagate the state multiple lengths of the crystal. In theory, it is possible to repeat this process as many times as needed, resulting in theory in an effective infinite nonlinear medium. In practice, this is not true, since the total number of passages may be bounded either by absorption or noise introduced during each reconstruction, which we estimated to be well below 7\% at each reinjection. Focusing on two case studies, we report a good qualitative agreement between experimental and numerical results up to $6$ propagation lengths, observing dynamics that were otherwise inaccessible while also allowing us to gather more information regarding the intermediate evolution of states. For the photorefractive crystal utilized, the anisotropic response may enclose additional dynamics that are not accounted for in the isotropic propagation model and deviate the obtained results for larger propagation distances. Yet we foresee that future implementations with isotropic mediums, such as hot atomic vapors or thermo-optical media, may bypass this limitation, allowing to establish successful analogue simulations of two-dimensional quantum fluids. 

Overall, we believe that the results enclosed offer an exciting path for a broad range of subsequent experimental works by enabling the experimental emulation of longer propagation distances in conjugation with additional access to the intermediate states.  In particular, combined with the capacity to generate states with arbitrary amplitude and phase profiles, it may set an important cornerstone for the exploration of topics such as two-dimensional quantum turbulence \cite{panico2023onset}, vortex dynamics in free and confined situations \cite{kwon2021sound}, or even analogue gravity studies \cite{braidotti2022measurement}.

\begin{acknowledgments}
This work is financed by National Funds through the Portuguese funding agency, FCT – Fundação para a Ciência e a Tecnologia, within project UIDB/50014/2020. T.D.F. is supported by Fundação para a Ciência e a Tecnologia through Grant No. SFRH/BD/145119/2019.
\end{acknowledgments}

\bibliography{feedbackloop}% Produces the bibliography via BibTeX.

\end{document}

\preprint{APS}

\title{Supplementary Material: Optical Feedback Loop in Paraxial Fluids of Light: A Gate to new phenomena in analogue physical simulations}
% Force line breaks with \\

\author{Tiago D. Ferreira}
\author{Ariel Guerreiro}
\author{Nuno A. Silva}
\address{Departamento de Física e Astronomia, Faculdade de Ciências, Universidade do Porto, Rua do Campo Alegre s/n, 4169-007 Porto, Portugal}
\address{INESC TEC, Centre of Applied Photonics, Rua do Campo Alegre 687, 4169-007 Porto, Portugal}

\date{\today}% It is always \today, today,

\maketitle

This document serves as the supplementary material for the manuscript entitled "Optical Feedback Loop in Paraxial Fluids of Light: A Gate to new phenomena in analogue physical simulations" and presents further details and considerations on the experimental setup, as well as, deeper insights on the experimental results presented in the main text.

%%%%%%%%%%%%%%%%%%%%%%%%%%%%%%%%%%%%%%%%%%%%%%%%%%%%%%%%%%%%%%%
%%%%%%%%%%%%%%%%%%%%%%%%%%%%%%%%%%%%%%%%%%%%%%%%%%%%%%%%%%%%%%%
%%%%%%%%%%%%%%%%%%%%%%%%%%%%%%%%%%%%%%%%%%%%%%%%%%%%%%%%%%%%%%%
\section{Paraxial Fluids of Light and the Hydrodynamic interpretation}

To develop a theoretical framework we start by considering that the optical beam is given by 
\begin{eqnarray}
\boldsymbol{E}_f=E_f(\boldsymbol{r}_\perp,z)\exp{[i (n_e\boldsymbol{k}_f z- \omega t)]}\boldsymbol{e}_p,
\label{eq:field_eq}
\end{eqnarray}
being the vacuum wavenumber $k_f = 2\pi / \lambda_f= \omega/(c)$ with $\lambda_{f}=532nm$, $c$ the velocity of light in vacuum \cite{ferreira2022towards}. We further consider that the $c-$axis of the photorefractive crystal is aligned with the polarization vector $\boldsymbol{e}_p$ (thus relating $n_e$ with the extraordinary refractive index). Assuming an isotropic response of the crystal and the paraxial approximation, the envelope dynamics may be described by the following equation
\begin{eqnarray}
i \frac{\partial E_f}{\partial z } + \frac{1}{2n_ek_f} \nabla_\perp^2 E_f + k_f\Delta n E_f + i\frac{\alpha}{2} E_f& = 0,
\label{eq:fluid_eq}
\end{eqnarray}
where $\alpha$ denotes the absorption. In the defocusing regime, the saturable nonlinear refractive index variation for a photorefractive crystal in the steady-state \cite{eloy2021experimental, boughdad2020fluids} may be written as
\begin{eqnarray}
    \Delta n=-\Delta n_{max} \frac{I_f }{I_f+ I_{sat}},
\end{eqnarray}
where $I_f=\left|E_f\right|^2$ is the local beam intensity, $I_{sat}$ an additional saturation intensity of the crystal that may be provided by an external white light, and $\Delta n_{max}=1/2n_e^3r_{33}E_{ext}$. 

The fluid of light analogy may be introduced by considering the Madelung transformation,
\begin{eqnarray}
\nonumber
\psi_f = \sqrt{I_f(\boldsymbol{r}_{\perp}, z)} e^{i \phi(\boldsymbol{r}_{\perp}, z)} 
\end{eqnarray}
where $\phi$ is the light fluid phase profile. In this new formalism, the nonlinear Scrhödinger equation may be rewritten as
\begin{eqnarray}
%parte imaginária
\frac{\partial I_f}{\partial z} + \nabla_\perp \cdot \left( I_f \boldsymbol{v} \right) = -\alpha I_f,
\label{eq:hydro_1}
\end{eqnarray}
\begin{eqnarray}
%parte real
\frac{\partial \boldsymbol{v}}{\partial z} + \left( \boldsymbol{v} \cdot \nabla_\perp\right) \boldsymbol{v} = \nabla_\perp\left(\frac{\Delta n}{n_e} + \frac{\nabla_\perp^2 \sqrt{I_f}}{2(n_e k_f)^2 \sqrt{I_f}}\right).
\label{eq:hydro_2}
\end{eqnarray}
In this form, neglecting both the absorption and the Bohm potential (the final terms on the right side of equation \ref{eq:hydro_1} and \ref{eq:hydro_2}, respectively), a direct comparison between light and fluid can be established via the definition of the fluid velocity as $\boldsymbol{v} = \nabla_\perp \phi/k_fn_e$, and by linking the local intensity to the fluid density. Furthermore, assuming that we are far away from the saturation, i.e. $\Delta n \approx I_f / I_{sat}$, one may obtain the Bogoliubov dispersion relation for the small excitations on top of a background fluid of density $I_f = I_0$ as
\begin{eqnarray}
\omega_z=\sqrt{\frac{k_{\perp}^2}{2k_fn_e}\left(\frac{k_{\perp}^2}{2k_fn_e}+2k_f\Delta n\right)}. 
\label{eq:bogolioubov}
\end{eqnarray}
Considering the low $k$ value, it follows the analogue sound velocity of the fluid
\begin{eqnarray}
c_s=\sqrt{\frac{\Delta n_{max} I_0}{n_e I_{sat}}}, 
\label{eq:sound_v}
\end{eqnarray}
and typical length scales in the transversal plane may be defined in terms of the healing length $\xi$, given by
\begin{eqnarray}
\xi=\frac{1}{k_f}\sqrt{\frac{I_{sat}}{\Delta n_{max} n_e I_0}}.
\label{eq:healing}
\end{eqnarray}

%%%%%%%%%%%%%%%%%%%%%%%%%%%%%%%%%%%%%%%%%%%%%%%%%%%%%%%%%%%%%%%
%%%%%%%%%%%%%%%%%%%%%%%%%%%%%%%%%%%%%%%%%%%%%%%%%%%%%%%%%%%%%%%
%%%%%%%%%%%%%%%%%%%%%%%%%%%%%%%%%%%%%%%%%%%%%%%%%%%%%%%%%%%%%%%

\section{Experimental Setup}
\begin{figure*}
\begin{center}
\includegraphics[width=1.0\textwidth]{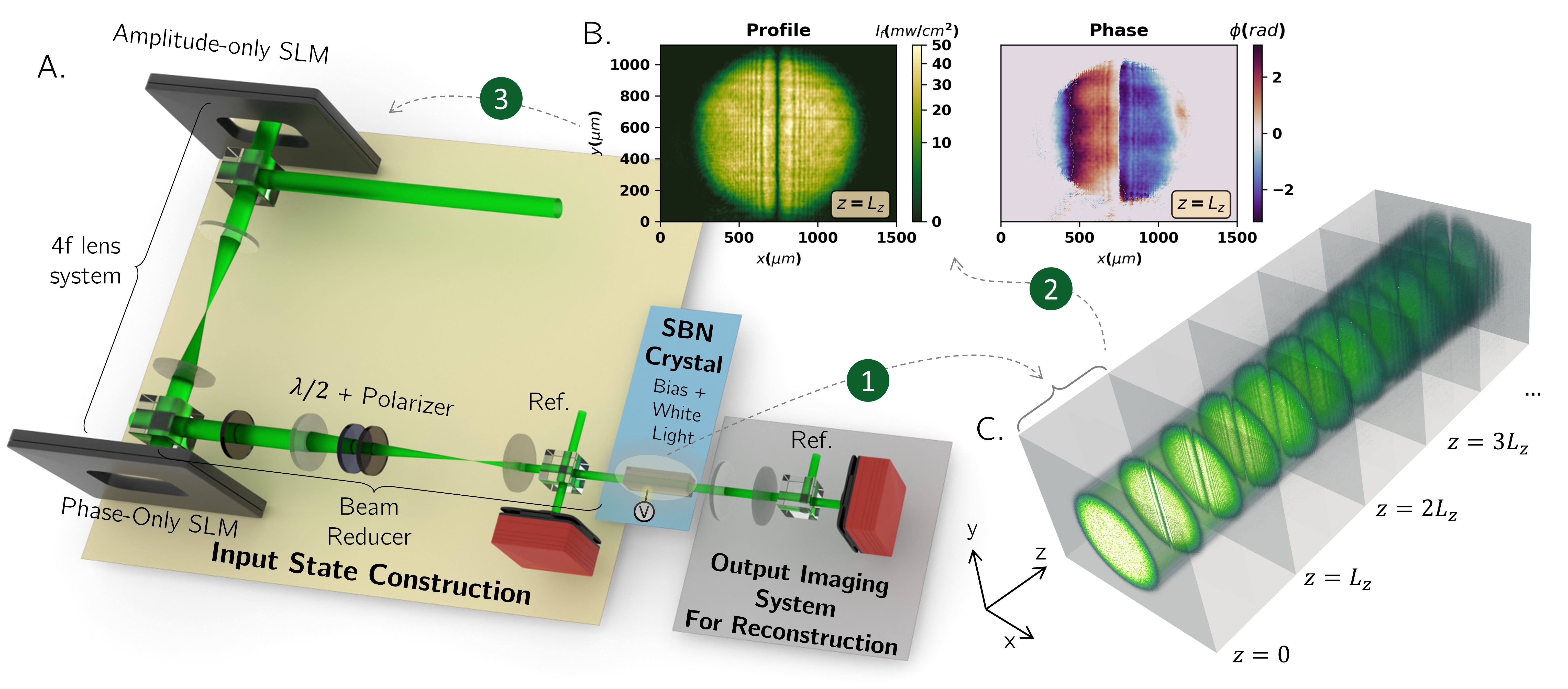}
\end{center}
\caption{Simplified illustration of the experimental setup, exploiting an SLM as an amplitude modulator matched at the single-pixel level to a second SLM acting as a phase-only modulator A. A reference arm is used for reconstructing the optical beam phase at the output face of the crystal, and Figure B. shows typical intensity and phase profiles obtained with the setup. Figure C. is a 3D representation of the evolution of the intensity with several passages through the crystal. The optical feedback loop represented in this schematic is composed of three fundamental steps (Discussed in more detail in this supplementary material). 1 - the state propagates through the crystal, and the intensity and phase at the output are retrieved - corresponding to 2. 3 - The output state is then reconstructed again at the entrance of the crystal for propagating an additional crystal length.}
\label{fig:experimental_setup}
\end{figure*}

\begin{figure*}
\begin{center}
\includegraphics[width=0.8\textwidth]{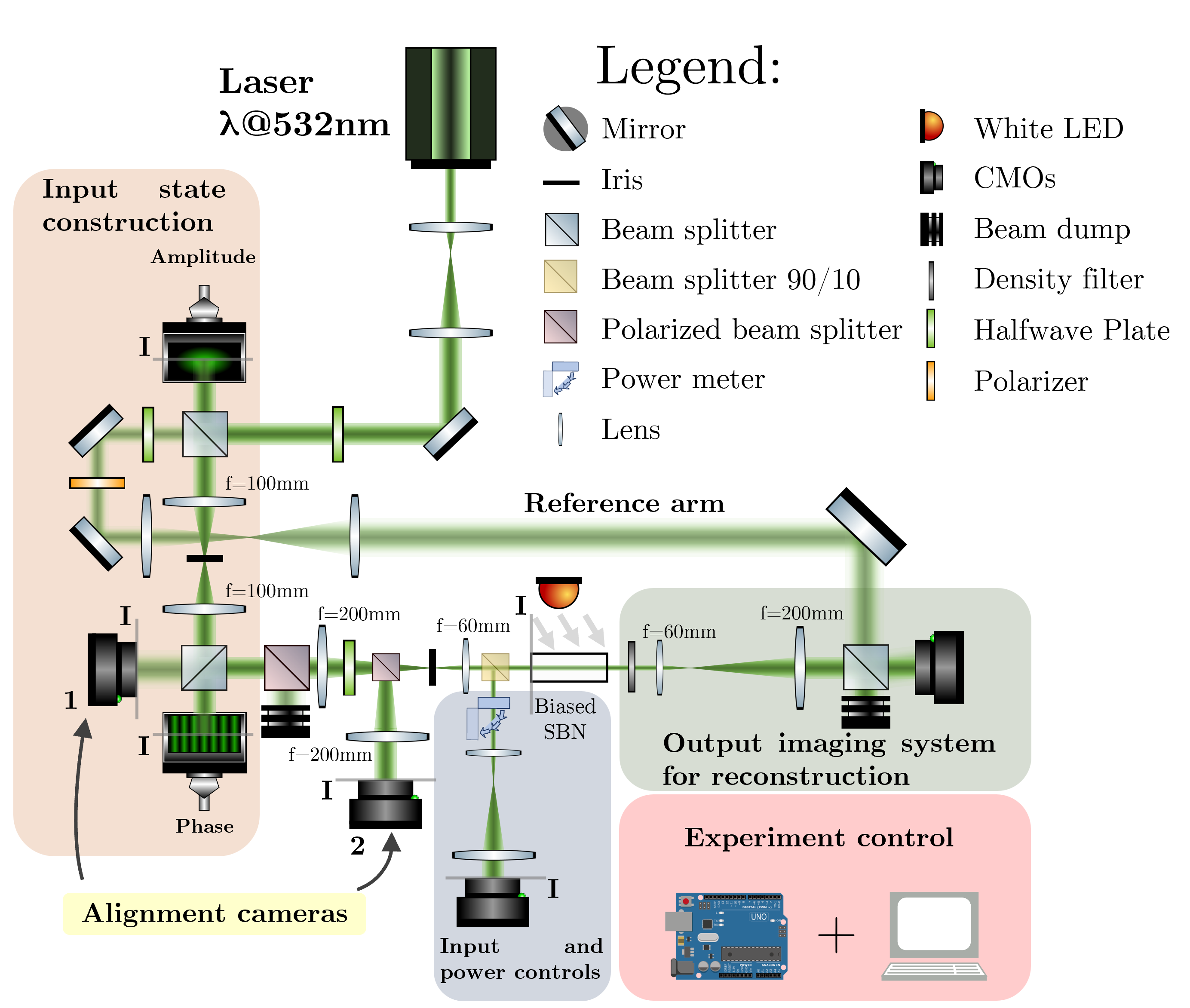}
\end{center}
\caption{Feedback loop complete experimental setup. A laser beam at $\lambda=532nm$ is expanded and collimated before entering the beam-shaping region. Here, amplitude-only and phase-only SLMs modulate the amplitude and phase profiles of the input field, respectively. The amplitude SLM image plane is imaged by a 4f lens system on the phase SLM image plane, and since both SLMs are similar, we have an alignment of the order of SLM pixel pith $8\mu m$. The input field image plane is then imaged at the entrance of the crystal through a 4f lens system. The gray lines across the experiment and with \textbf{I} close to it, indicate the experimental positions that have the same image plane. We use a motorized halfwave plate to adjust the input power. An additional 4f lens system images the output plane, and a reference arm is used to obtain the interferograms and reconstruct the phase profile. The experiment is fully automated and controlled by an Arduino device.}
\label{fig:experimental_setup_complete}
\end{figure*}

In Figure \ref{fig:experimental_setup}, we present an expanded version of Figure 1 of the main text for visibility purposes. In Figure \ref{fig:experimental_setup_complete}, we have a detailed version of the experimental setup used in this work. A laser beam at $\lambda=532nm$ is expanded and enters the beam shaping area. In this part of the setup, amplitude-only and phase-only SLMs are responsible for modulating the amplitude and phase profiles of the input beam, respectively. This is accomplished by projecting the first SLM image plane onto the second SLM image plane using a 4-f lens system with a magnification factor of $1$. This allows us to have a match between the two SLMs with a pixel pitch precision, i.e. $8\mu m$. The complete modulated field is then imaged at the entrance of the crystal through another 4f lens system, but now with a magnification factor $60/200$. For alignment purposes, we have two auxiliary cameras. The first one is positioned at the image plane of the first SLM to align the crystal entrance with the beam-shaping region. The second alignment camera allows to match both SLMs masks at single-pixel level. The input power is controlled in each passage using a motorized half-wave plate combined with a polarizing beam splitter and monitored with a power meter, which allows an accuracy below $1\mu W$. Before the crystal, we also have an input imaging system that allows us to control the input state and correct for the Gaussian shape of the input beam (discussed further). A final 4f system is responsible for imaging the output of the crystal, and a reference arm is used to create the interferograms and retrieve the phase profile using digital off-axis digital holography technique \cite{OADH_method}. The experimental setup is fully automated and controlled through an Arduino device and a dedicated Python library developed in specific for this experimental setup. This allows the creation of arbitrary input conditions and permits remote operation of the experimental setup.

%%%%%%%%%%%%%%%%%%%%%%%%%%%%%%%%%%%%%%%%%%%%%%%%%%%%%%%%%%%%%%%
%%%%%%%%%%%%%%%%%%%%%%%%%%%%%%%%%%%%%%%%%%%%%%%%%%%%%%%%%%%%%%%
%%%%%%%%%%%%%%%%%%%%%%%%%%%%%%%%%%%%%%%%%%%%%%%%%%%%%%%%%%%%%%%

%\begin{figure}
%\begin{center}
%\includegraphics[width=1.0\textwidth]{feedback_scheme.jpg}
%\end{center}
%\caption{Ar.}
%\label{fig:feedback_scheme}
%\end{figure}

%%%%%%%%%%%%%%%%%%%%%%%%%%%%%%%%%%%%%%%%%%%%%%%%%%%%%%%%%%%%%%%
%%%%%%%%%%%%%%%%%%%%%%%%%%%%%%%%%%%%%%%%%%%%%%%%%%%%%%%%%%%%%%%
%%%%%%%%%%%%%%%%%%%%%%%%%%%%%%%%%%%%%%%%%%%%%%%%%%%%%%%%%%%%%%%

\section{The Optical Feedback Loop}
Our implementation of the optical feedback loop, schematically represented in Figures \ref{fig:experimental_setup} and \ref{fig:experimental_setup_complete}, is initialized by creating a state 
\begin{equation}
    E_{in}(x, y; i=0)=A_{in}(x, y; i=0)e^{i\phi_{in}(x, y; i=0)},
\end{equation}
 where $|A_{in}(x, y; i=0)|^2$ and $\phi_{in}(x, y; i=0)$ are the input intensity and phase profiles, respectively, and the index $i$ tracks the passage number inside the crystal. The optical feedback loop proceeds as following:

 \begin{enumerate}
 
  \item The first step corrects the Gaussian shape of the input beam. We utilize the input camera to measure the Gaussian shape $f(x,y)$. Then, to modulate an arbitrary amplitude profile, let us call it $A_{in}(x,y)$, it is necessary to divide it by the Gaussian shape, as
    \begin{equation}
        |A_{in}^{corr}(x, y)|^2 \propto \frac{|A_{in}(x,y)|^2}{f(x,y)},
    \end{equation}
    where $|A_{in}^{corr}|^2$ is the corrected amplitude profile to be passed to the amplitude modulator. It should be noted that in the absence of this adjustment, we would multiply the intended state by $f(x,y)$ in each passage, which would distort the amplitude profile and produce incorrect results. 
 
  \item After measuring the Gaussian profile of the input beam, $f(x,y)$, as discussed in the previous point, the amplitude $|A_{in}^{corr}(x, y; i=0)|^2$ is applied to the amplitude-only SLM, and $\phi_{in}(x, y; i=0)$ to phase-only SLM. Its input power is adjusted to the desired value by rotating the half-wave plate accordingly. 
  
  \item The state then propagates through the crystal, and the output is measured by the imaging system. Here, a CMOS camera records the intensity, and the phase is retrieved through the digital off-axis holographic method. Before being reinjected again into the system, a Gaussian filter with a kernel of $10$ pixels of size is applied to the intensity profile to reduce some noise and artifacts introduced by the nonlinear response of the crystal and some dust in the setup. 

  \item The state is then ready to enter the system again, with the new amplitude and phase profiles, $|A_{in}(x, y; i=1)|^2/f(x,y)$ and $\phi_{in}(x, y; i=1)$, respectively, applied to their respective SLM. The following power adjustments take into account that the crystal absorbs some of the light during each passage, which was determined beforehand.
  
\end{enumerate}

The process is automatized and performed as many times as desired. 

%%%%%%%%%%%%%%%%%%%%%%%%%%%%%%%%%%%%%%%%%%%%%%%%%%%%%%%%%%%%%%%
%%%%%%%%%%%%%%%%%%%%%%%%%%%%%%%%%%%%%%%%%%%%%%%%%%%%%%%%%%%%%%%
%%%%%%%%%%%%%%%%%%%%%%%%%%%%%%%%%%%%%%%%%%%%%%%%%%%%%%%%%%%%%%%

\section{Comment on the Quality of Arbitrary State Generation}

\begin{figure}[!htb]
\begin{center}
\includegraphics[width=0.5\textwidth]{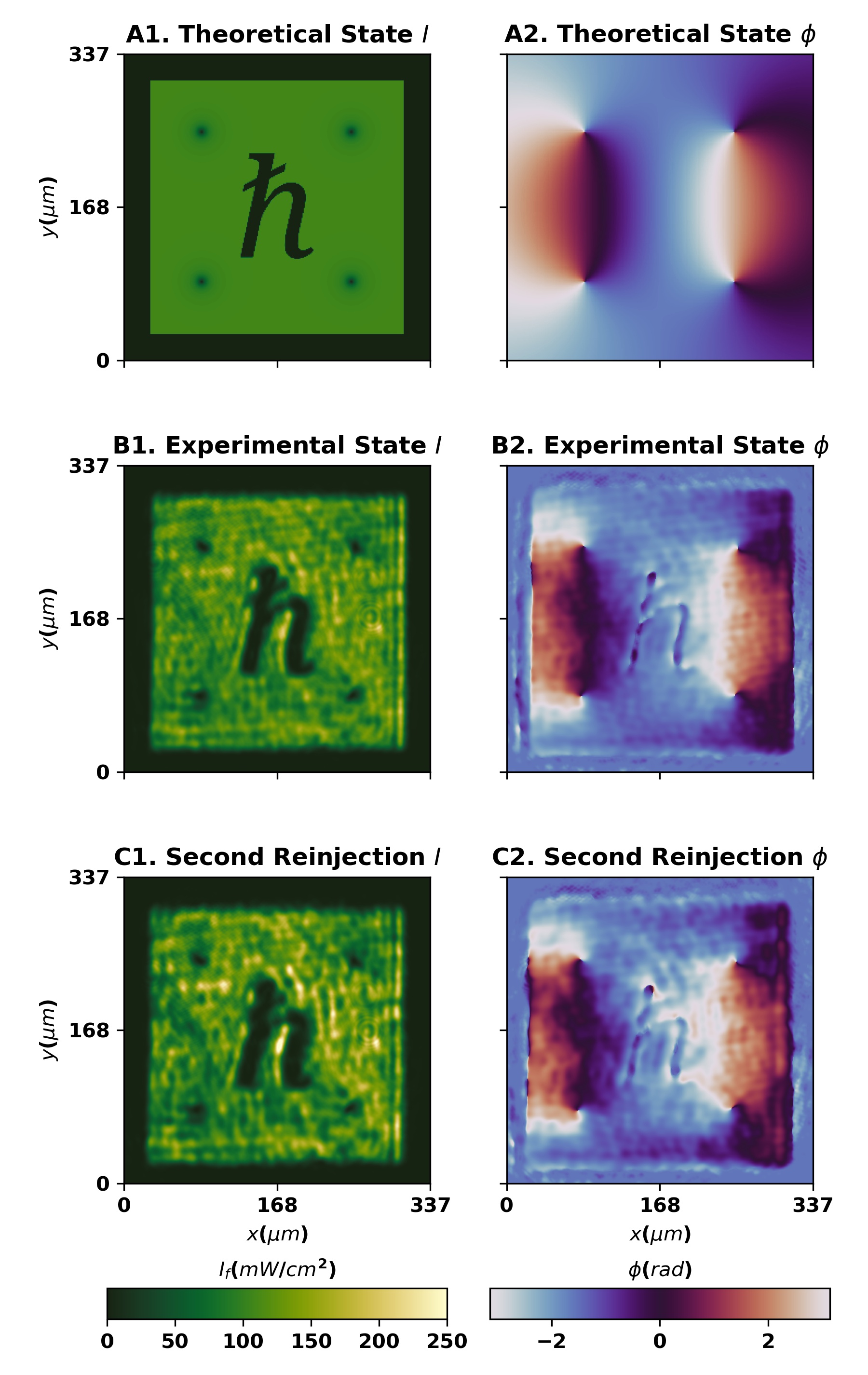}
\end{center}
\caption{Arbitrary state generation. In Figure A we have the numerical state, and in B the measured experimental state. This experimental state is then reinjected into the system, and the corresponding output is in Figure C. These results are obtained by imaging the crystal input image plane.}
\label{fig:arbitrary_state_generation}
\end{figure}

As presented in the main manuscript, the setup allows to create arbitrary states in amplitude and phase. To showcase this ability and validate the methodology, Figure \ref{fig:arbitrary_state_generation} depicts a challenging amplitude profile with sharp edges and vortex phase profiles. Measuring the state at the input, we may evaluate in terms of the Absolute Percentage Error of the amplitude as defined in the main text, 
\begin{equation}
APE (|A_{in,exp}(x,y)|) = 100 \times \left| 1- \frac{|A_{in,exp}(x,y)|}{|A_{in}(x,y)|}\right|\%,
\label{eq:MAPE_amp_supp}
\end{equation}
which gives a mean absolute percentage error below 6\% for a first reconstruction and below 10\% for a second reconstruction, i.e. reconstructing previously measured state at the input plane (computed considering regions of $I>1mW/cm^{2}$. Although additional noise on the state on the second reconstruction increases as expected, the overall profile and phase are still preserved.

\begin{figure}[!htb]
\begin{center}
\includegraphics[width=0.5\textwidth]{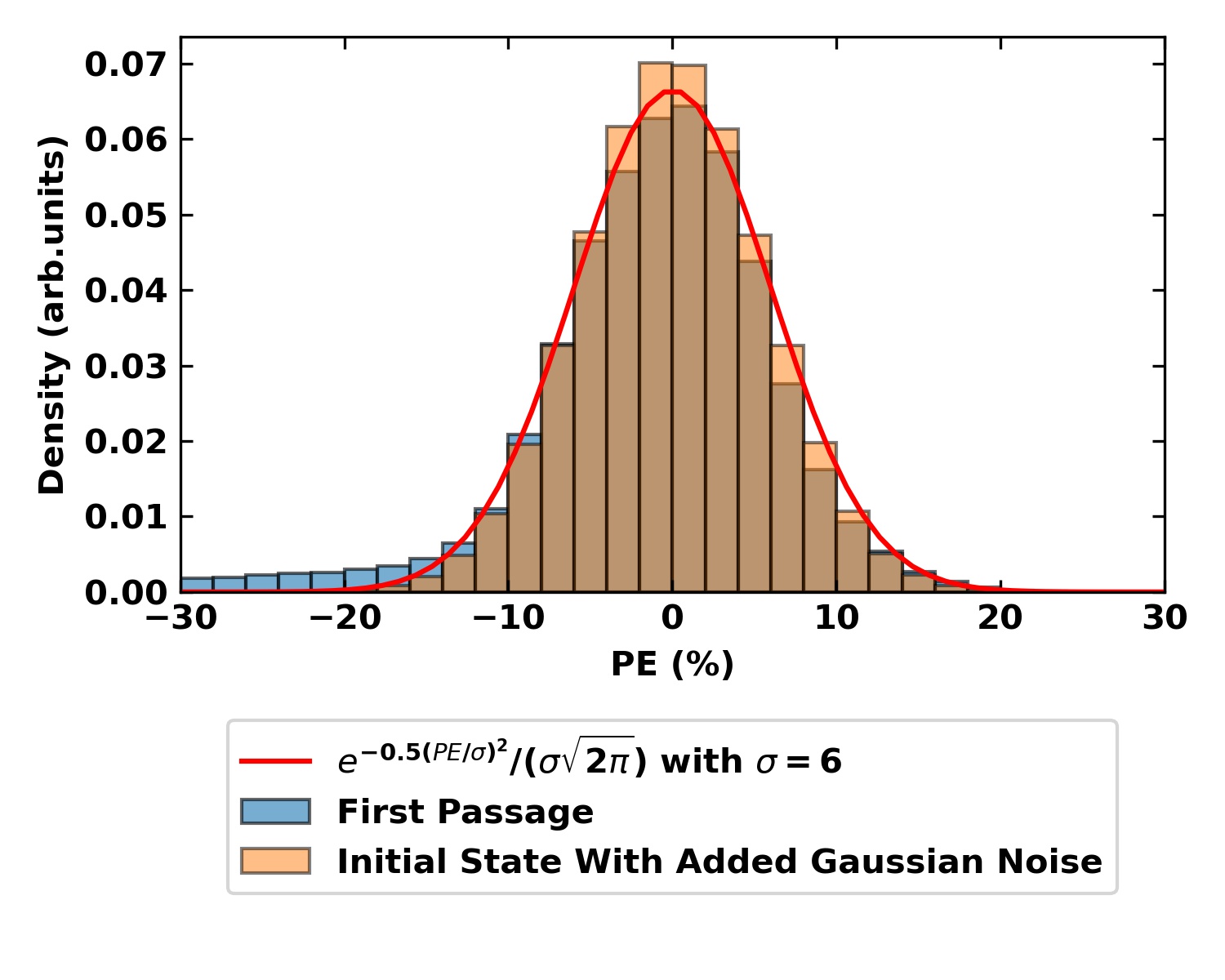}
\end{center}
\caption{Noise distribution of the created state compared with the theoretical one, represented by the blue bars. This compassion is for the complex state. The error distribution follows closely a Gaussian distribution, indicated by the red line. Indeed, if we add some Gaussian noise to the theoretical state and perform a similar analysis, we get a similar noise distribution, represented by the orange bars.}
\label{fig:error}
\end{figure}

To further infer the type of noise created by the reconstruction process at each passage and to provide a good estimate to the noise added to the numerical simulations, Figure \ref{fig:error} presents the percentage error distribution for the measured complex state compared with a theoretical one. The results show that the error closely follows a Gaussian distribution with a $\sigma=6$, meaning that it can be well approximated with white gaussian noise at numerical level and as further demonstrated in Figure \ref{fig:error}.

There are some factors that we have to take into account that help to explain these errors:
\begin{itemize}

    \item \textbf{Dust} - The principal source of noise is the dust along the beam path, more susceptible to appearing due to the higher Rayleigh scattering at the visible wavelengths when compared to an infrared laser beam. It is also possible to observe that the experimental noise seems to be slightly shifted towards negative values, which may be related to a systematic loss of power along the beam path.

    \item \textbf{Low powers} - With the photorefractive crystal, we need low powers, $\sim10-100\mu W$, to avoid saturating its nonlinear response. Consequently, this reduces the signal-to-noise ratio, presenting a more challenging scenario when compared with atomic vapors, for example.
    
    \item \textbf{Large magnifications} - In the beam-shaping stage, we aim to use the largest Gaussian profile possible to increase the resolution available, thus requiring a beam reducer lens system with a large magnification factor before entering the crystal. This not only adds further complexity to the alignment process but also may introduce some aberrations in the wavefront that are difficult to correct with the current experimental configuration as the input camera is not at the crystal plane but rather at a conjugated image plane.
    
    \item \textbf{SLMs Bit Depth and resolution} - In this experimental setup, we use two SLMs (HOLOEYE - PLUTO-VIS-016 for the phase modulation, HOLOEYE - HES 7020 6001 for the amplitude modulation) with the same pixel pitch, $8\mu m$, and both can modulate up to $256$ gray levels. This $8$-bit codification, limits the capacity of reconstructing the states, since small intensity and phase modulations may be erased when we send the profiles to the Spatial Light Modulators due to the limited dynamic range and resolution. SLMs with higher bit depth are scarce and more expensive, but may present a way to circumvent this limitation for experiments with higher requirements.
    
\end{itemize}

%%%%%%%%%%%%%%%%%%%%%%%%%%%%%%%%%%%%%%%%%%%%%%%%%%%%%%%%%%%%%%%
%%%%%%%%%%%%%%%%%%%%%%%%%%%%%%%%%%%%%%%%%%%%%%%%%%%%%%%%%%%%%%%
%%%%%%%%%%%%%%%%%%%%%%%%%%%%%%%%%%%%%%%%%%%%%%%%%%%%%%%%%%%%%%%

\section{On the effects of noise and importance of the power correction}

Our entire approach is based on the assumption that we inject the same optical state obtained at the end plane of the crystal, 
\begin{equation}
    E_{in}(x, y; i=1)=A_{out}(x, y; i=1)e^{i\phi_{out}(x, y; i=1)},
\end{equation}
Yet, as the system is not ideal, we are reinjecting with an additional term
\begin{equation}
    E_{in}(x, y; i=1)=A_{out}(x, y; i=1)e^{i\phi_{out}(x, y; i=1)} + \delta(x,y,z),
\end{equation}
representing the noise/deviation of the output state. This will evolve nonlinearly inside the crystal creating a nonlinear deviation from the expected solution at the end of the next passage. Indeed, note that the same effect also happens at the very first passage seeded by background noise on top of the initial state, and has been investigated in previous works \cite{steinhauer2022analogue}.

To discuss the consequences of this we shall note that when the optical beam enters the crystal, it feels a sudden change in the interaction strength provided by the nonlinear optical response. As the nonlinear refractive index rapidly changes from $0$ to $\Delta n (I)$, the state undergoes effects that resemble quantum quenches on Bose-Einstein Condensates. From this point on, the evolution will follow the NSE, and a possible approach to address the effects of such noise is to consider that the noise is perturbative, as in the case discussed for obtaining the hydrodynamic interpretation. Substituting directly in the NSE and linearizing it yields the equation system
\begin{equation}
i\partial_{z}\left(\begin{array}{c}
\delta\\
\delta^{*}
\end{array}\right)=\left(\begin{array}{cc}
D_{K} & k_{f}\Delta n_{max}\left(E_{in}\right)^{2}\\
-k_{f}\Delta n_{max}\left(E_{in}^{*}\right)^{2} & -D_{K}
\end{array}\right)\left(\begin{array}{c}
\delta\\
\delta^{*}
\end{array}\right),
\end{equation}
where 
\begin{equation}
D_{K}=-\frac{1}{2n_{e}k_{f}}\nabla^{2}+2k_{f}\Delta n_{max}\left|E_{in}\right|^{2}.
\end{equation}
However, note that in this case we no longer have a constant fluid density or constant phase, meaning that the Fourier approach is no longer straightforward. Typically, addressing the linearized equation system for small excitations from a purely analytical perspective becomes challenging and can only be performed with numeric approaches. Nevertheless, as it happens for the simplest case, the input beam will still be decomposed into the eigenmodes of the Bogoliubov problem. In practice, this translates into a decomposition of the noise on top of the input field in the $k-$space, creating excitations with multiple values of $k$ and whose effect may be non-negligible due to interferences \cite{ferreira2018superfluidity,fontaine2018observation,fontaine2020interferences} or to the instability of some of the modes \cite{giacomelli2023interplay} and affect the overall state dynamics. In the particular context of possible unstable modes, reducing the propagation distance may work as a mitigation plan, but may introduce additional noise due to the higher number of passages and reconstructions.

\begin{figure}
\begin{center}
\includegraphics[width=1\textwidth]{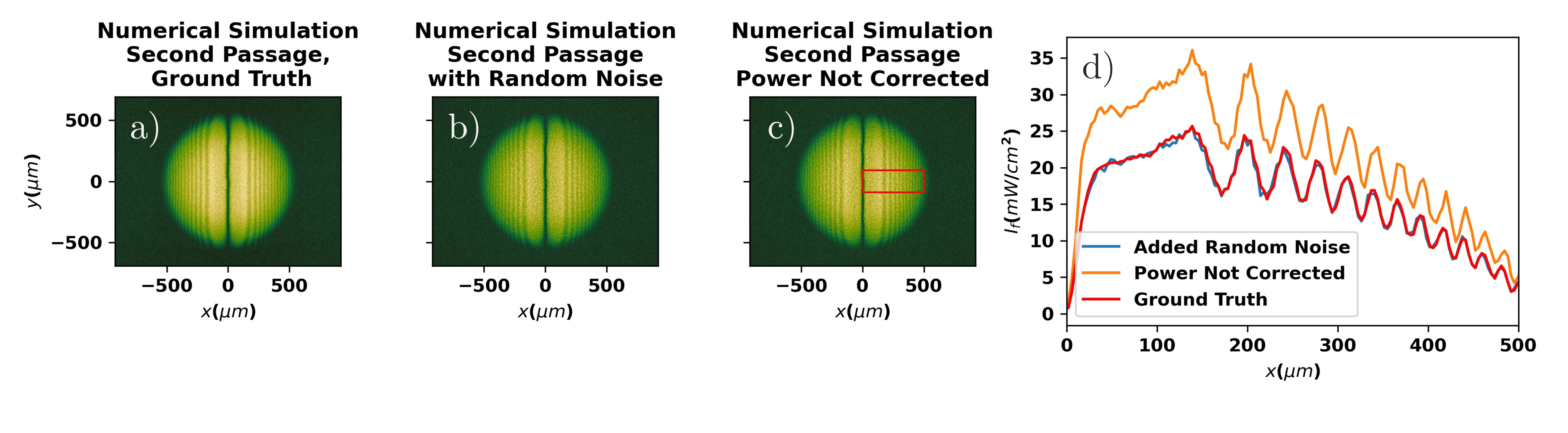}
\end{center}
\caption{Numerical results to test the effect of the noise and power correction on the second passage. a) ground truth after the second passage; b) when random noise is added after the first passage; c) when the power is not corrected after the first passage. Image d) shows a comparison between the profiles of a)-c). Note that while the random noise produces very small deviations from the expected dynamics, this is not true for the case where the power is not corrected, where some deviations are observed.}
\label{fig:quench}
\end{figure}

To better illustrate the appearance of additional perturbations we performed numerical simulations for the optical reinjection scenario, whose results are depicted in Figure \ref{fig:quench}. In particular, we focused on three distinct situations, i) reinjecting the measured state, ii) adding random noise for the second passage, and iii) injecting without correcting the power (i.e. injecting the same optical power as the initial solution, meaning that we neglect the power lost by absorption). Taking the first scenario as the ground truth, we see that the quench is minimal for the random noise case but leads to non-negligible deviations observed for the uncorrected power case in comparison with the expected solution. 

As the effects of the quench from the random noise seem to be negligible for this case study, it means that the initial assumption seems to be good enough for the cases evaluated in the manuscript, which is empirically confirmed by taking into account the experimental and numerical results presented throughout this work. Nevertheless, we believe that the study of the effects associated with this analogue quench in this experimental configuration presents an exciting research path for future works, not only from the experimental perspective by attempting to minimize their effect with better reconstruction and injection methodologies, but also from a theoretical perspective, utilizing this system as a playground for generating and studying quench in such systems \cite{steinhauer2022analogue}.

%%%%%%%%%%%%%%%%%%%%%%%%%%%%%%%%%%%%%%%%%%%%%%%%%%%%%%%%%%%%%%%
%%%%%%%%%%%%%%%%%%%%%%%%%%%%%%%%%%%%%%%%%%%%%%%%%%%%%%%%%%%%%%%
%%%%%%%%%%%%%%%%%%%%%%%%%%%%%%%%%%%%%%%%%%%%%%%%%%%%%%%%%%%%%%%

\section{Experimental Results}

\subsection{Dark Soliton Decay through Snake Instability}

\begin{figure}
\begin{center}
\includegraphics[width=0.5\textwidth]{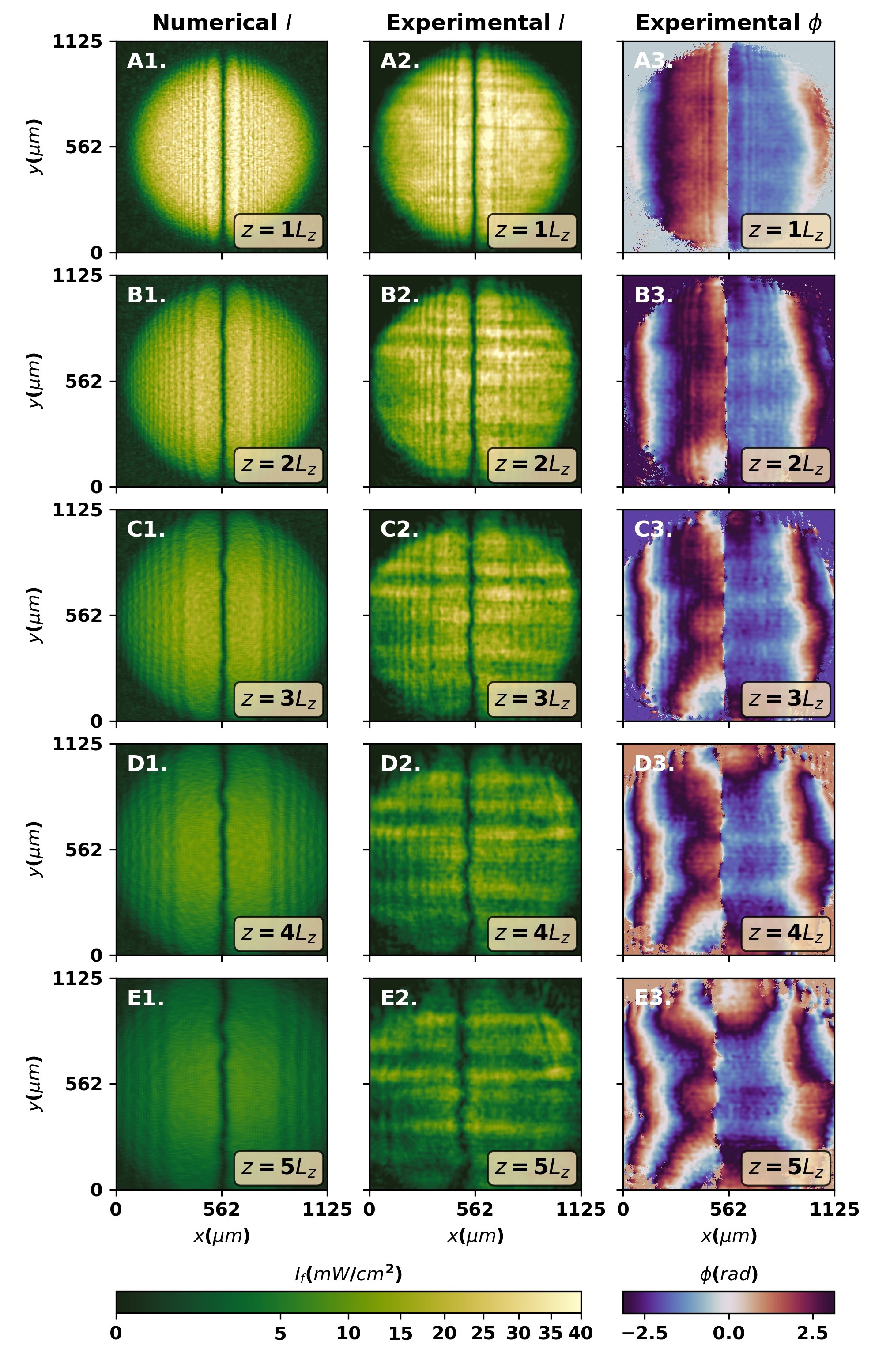}
\end{center}
\caption{Complete dark soliton decay, snake instability, and shock waves expansion. The first column corresponds to the numerical simulations, while the last two to experimental data, with the amplitude and phase profiles, corresponding to the middle and last columns, respectively. In the last row, a section of the system is highlighted to show the dark soliton decay into vortex pairs. Certain modulations caused by the crystal nonlinear response can be seen in the experimental data. In particular, due to the $E_{ext}$. These modulations are amplified during the propagation and influence the position where the vortices are formed.}
\label{fig:darksol_sup}
\end{figure}

In this section, we expand the results presented in the main manuscript for the Dark soliton decay depicting all the multiple passages measured in the experiment in Figure \ref{fig:darksol_sup}. Two phenomena can be compared when analyzing the experimental data: the shock wave expansion and snake instability that leads to the vortex formation. A good qualitative agreement is found in both parameters for all the passages.

%%%%%%%%%%%%%%%%%%%%%%%%%%%%%%%%%%%%%%%%%%%%%%%%%%%%%%%%%%%%%%%
%%%%%%%%%%%%%%%%%%%%%%%%%%%%%%%%%%%%%%%%%%%%%%%%%%%%%%%%%%%%%%%
%%%%%%%%%%%%%%%%%%%%%%%%%%%%%%%%%%%%%%%%%%%%%%%%%%%%%%%%%%%%%%%

\subsection{Collision of Flat-top states and vortex dynamics}

\begin{figure}
\begin{center}
\includegraphics[width=0.5\textwidth]{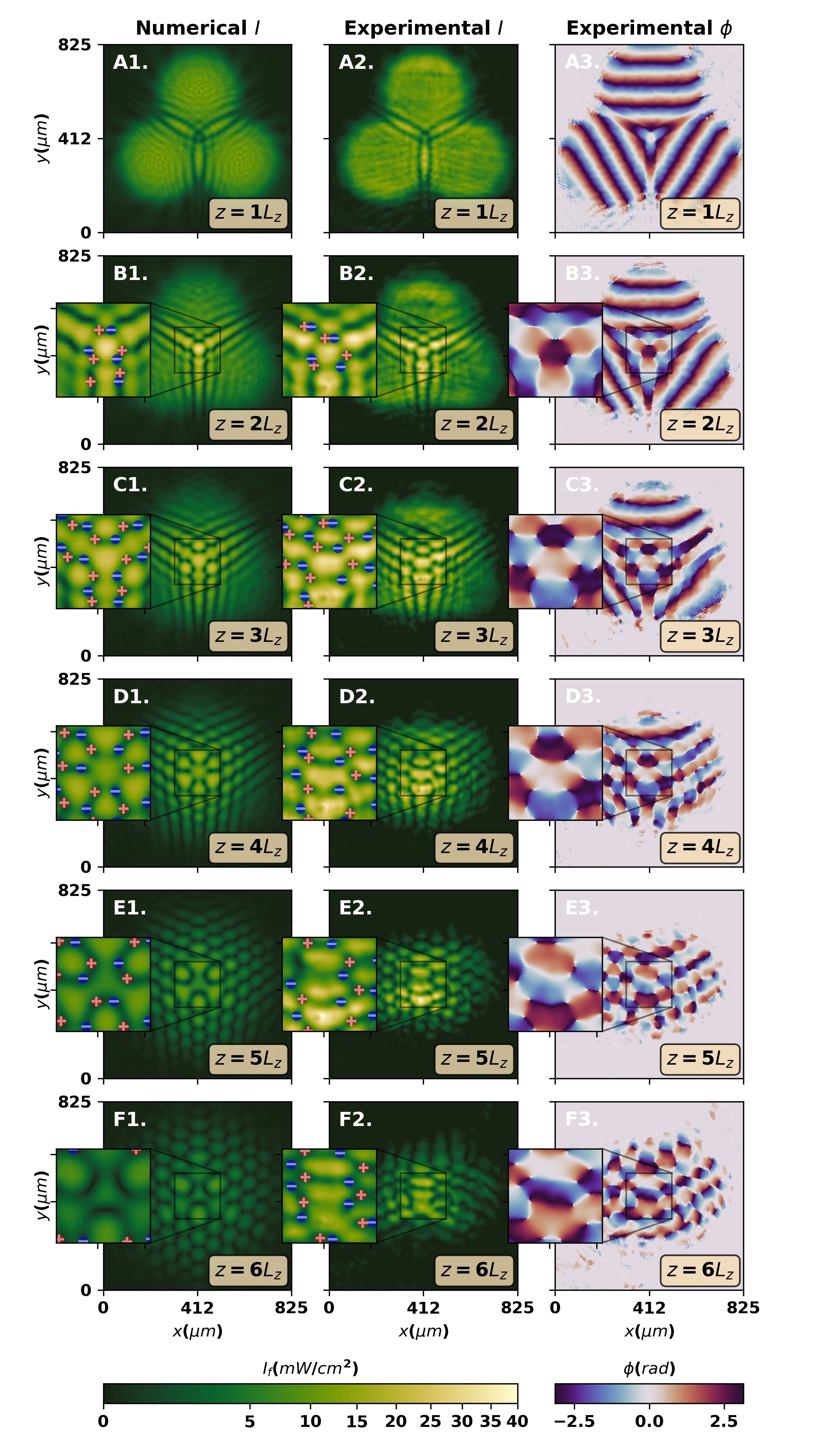}
\end{center}
\caption{Complete collision dynamics between three flat-top states. The first column corresponds to the numerical simulations, while the last two to experimental data, with the amplitude and phase profiles, corresponding to the middle and last columns, respectively. The center of the collision is highlighted to help the comparison between the numerical and experimental data. The red circles with a "$+$" and the blue circles with a "$-$" correspond to the vortices of positive and negative circulation, respectively.}
\label{fig:flatop_sup}
\end{figure}

In this section, we expand the results presented in the main manuscript for the collision between three flat-top states and the corresponding vortex dynamics, for all the passages measured in the experiment, Figure \ref{fig:flatop_sup}. As observed, up until the passage $4$, the numerical and experimental results agree quite well, where in both results it is possible to observe the formation of a symmetric structure composed of vortices. However, for the next passages, the anisotropic response of the crystal, not taken into account in the numerical solver, seems to start to influence the dynamics as it breaks the symmetry of the central structure. As it is possible to see in the passage from Figure \ref{fig:flatop_sup} - D2 and to Figure \ref{fig:flatop_sup} - E2, the central structure is stretched along the $x-$axis, which is parallel to the $c-$axis of the crystal. Contrary to the previous case study, the vortices in this configuration are less affected by the anisotropy of the crystal. This occurs because now the system nucleates many vortex pairs close to each other and with different orientations concerning the $c-$axis. This is known to help to stabilize their velocity field, and, consequently, their structure \cite{mamaev1996vortex,mamaev1997time}. In these last passages, it is more difficult to compare the experimental results with the numerical ones. Nevertheless, it is possible to track the dynamics of the vortices in the experimental results, which reinforces the wide range of non-trivial opportunities introduced by the implemented feedback loop. 

%%%%%%%%%%%%%%%%%%%%%%%%%%%%%%%%%%%%%%%%%%%%%%%%%%%%%%%%%%%%%%%
%%%%%%%%%%%%%%%%%%%%%%%%%%%%%%%%%%%%%%%%%%%%%%%%%%%%%%%%%%%%%%%
%%%%%%%%%%%%%%%%%%%%%%%%%%%%%%%%%%%%%%%%%%%%%%%%%%%%%%%%%%%%%%%

%%%%%%%%%%%%%%%%%%%%%%%%%%%%%%%%%%%%%%%%%%%%%%%%%%%%%%%%%%%%%%%
%%%%%%%%%%%%%%%%%%%%%%%%%%%%%%%%%%%%%%%%%%%%%%%%%%%%%%%%%%%%%%%
%%%%%%%%%%%%%%%%%%%%%%%%%%%%%%%%%%%%%%%%%%%%%%%%%%%%%%%%%%%%%%%

%%%%%%%%%%%%%%%%%%%%%%%%%%%%%%%%%%%%%%%%%%%%%%%%%%%%%%%%%%%%%%%
%%%%%%%%%%%%%%%%%%%%%%%%%%%%%%%%%%%%%%%%%%%%%%%%%%%%%%%%%%%%%%%
%%%%%%%%%%%%%%%%%%%%%%%%%%%%%%%%%%%%%%%%%%%%%%%%%%%%%%%%%%%%%%%

\bibliography{feedbackloop}% Produces the bibliography via BibTeX.